\documentclass[conference]{IEEEtran}    

\IEEEoverridecommandlockouts 

\usepackage[T1]{fontenc}
\usepackage[utf8]{inputenc}

\usepackage{amsmath,amssymb,amsfonts}

\usepackage{graphicx}
\usepackage{subcaption}

\usepackage{array,booktabs}

\usepackage{cite}

\usepackage{algorithmic}
\usepackage{listings}

\usepackage[
    bookmarks=true,
    breaklinks=true,
    letterpaper=true,
    colorlinks=true,
    citecolor=blue,
    linkcolor=blue,
    urlcolor=blue
]{hyperref}

\usepackage{xcolor}
\usepackage{enumitem}
\usepackage{url}
\usepackage[normalem]{ulem}
\usepackage[most]{tcolorbox}
\usepackage[all]{nowidow}
\usepackage{comment}
\usepackage{dirtree}

\usepackage{orcidlink}

\definecolor{darkgreen}{rgb}{0,0.5,0}

\def\BibTeX{{\rm B\kern-.05em{\sc i\kern-.025em b}\kern-.08em
    T\kern-.1667em\lower.7ex\hbox{E}\kern-.125emX}}

\newcommand{\eg}{\textit{e.g.}}
\newcommand{\ie}{\textit{i.e.}}
\newcommand{\etc}{\textit{etc}}

\newcommand{\as}{\textsf{AS-CPU}}
\newcommand{\ts}{\textsf{TS-CPU}}
\newcommand{\ooo}{\textsf{O3-CPU}}
\newcommand{\ruby}{\textsf{Ruby}}
\newcommand{\garnet}{\textsf{Garnet}}
\newcommand{\slicc}{\textsf{SLICC}}

\newcommand{\red}[1]{\textcolor{red}{#1}}
\newcommand{\orange}[1]{\textcolor{orange}{#1}}
\newcommand{\blue}[1]{\textcolor{blue}{#1}}
\newcommand{\purple}[1]{\textcolor{purple}{#1}}
\newcommand{\olive}[1]{\textcolor{olive}{#1}}
\newcommand{\teal}[1]{\textcolor{teal}{#1}}

\newcounter{observation}
\newcommand{\nextobservation}{
\stepcounter{observation}
\textbf{~Observation~\arabic{observation}.}
}

\newcounter{summary}
\newcommand{\nextsummary}{
\stepcounter{summary}
\textbf{~Summary~\arabic{summary}.}
}

\usepackage{eso-pic}

\begin{document}
\IEEEpubid{%
\parbox{\columnwidth}{\centering
Published in the Proceedings of the 2026 IEEE International Symposium on Performance Analysis of Systems and Software (ISPASS)
\hfill
}
\hspace{\columnsep}\makebox[\columnwidth]{}
}
\vspace{0cm}
\AddToShipoutPictureFG*{%
\AtPageUpperLeft{%
\hspace*{\dimexpr0.5\paperwidth\relax}
\raisebox{-1.5cm}{%
\centering
    \includegraphics[width=9cm]
    {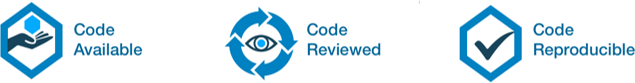}%
}}}

\title{Understanding Simulated Architecture via \\ gem5 Call-Stack Profiling}

\author{\IEEEauthorblockN{Johan S\"oderstr\"om\orcidlink{0009-0004-5420-3118}}
\IEEEauthorblockA{johan.soderstrom.9461@student.uu.se \\
\textit{Uppsala University}\\
Uppsala, Sweden}
\and
\IEEEauthorblockN{Rashid Aligholipour\orcidlink{0000-0001-9376-2925}}
\IEEEauthorblockA{rashid.aligholipour@it.uu.se \\
\textit{Uppsala University}\\
Uppsala, Sweden}
\and
\IEEEauthorblockN{Yuan Yao\orcidlink{0000-0001-9448-5595}}
\IEEEauthorblockA{yuan.yao@it.uu.se \\
\textit{Uppsala University}\\
Uppsala, Sweden}
}

\maketitle
\pagestyle{empty}
\pagenumbering{arabic}

\begin{abstract}

Understanding the behavior of simulated architectures in gem5 is critical for studying complex, deeply integrated computing systems. However, conventional analysis methods, which rely heavily on simulation statistics, provide only an indirect view of the simulated system internals. In this work, we show that call-stack profiling of gem5 itself offers a powerful yet underutilized perspective: the simulator's own call-stack directly reflects the activity of the simulated system, exposing insights that conventional statistics may overlook.

Profiling gem5's call-stacks, however, is challenging due to its highly layered and complex software design patterns. To address this, we introduce a specialized, lightweight profiling framework built on Linux's \texttt{perf\_event} interface which samples and analyzes gem5's runtime call-stacks throughout the simulation, resolves symbols on the fly, and merges samples into a hierarchical call-tree representation supporting both high-level structural views and focused, user-defined, component-specific analysis. Moreover, all profiling is performed in a dedicated helper process running alongside the main gem5 process, avoiding intrusive changes and overheads to the simulation itself.

We apply our framework to gem5's three major CPU models---AtomicSimpleCPU, TimingSimpleCPU, and O3CPU---together with the Ruby memory system, and uncover behaviors that are not easily observable in conventional gem5 statistics. Our case studies reveal, for example, that TimingSimpleCPU is inefficient due to its use of a lockup-cache model and, despite its conceptual simplicity, does not simulate faster than a full out-of-order core. In addition, our tool makes it straightforward to detect cache coherence protocol deadlock and livelock---issues that are otherwise difficult to identify, since the simulation either appears to run normally or terminates abruptly, making it hard to pinpoint when these conditions occur.

\end{abstract}

\section{Introduction}

\textbf{Problem.} The gem5 simulator~\cite{binkert2011gem5, lowe2020gem5} is widely used in computer architecture research for its configurability and flexibility in modeling modern computing systems. However, despite its broad adoption, a practical yet fundamental challenge remains: understanding exactly what the simulator is doing. As architectural complexity grows---particularly with innovative many-core architectures and deep memory hierarchies---gem5's internals become increasingly intricate and difficult to reason about, making it challenging to identify simulation bottlenecks, examine microarchitectural dynamics, or detect performance bugs.

\textbf{Prior work} has examined gem5 primarily from the perspective of host-machine resource usage (\eg, branch prediction accuracy, cache miss rates, memory footprint)~\cite{umeike2023profiling}, or leverages profiling data---specifically hardware and simulated performance counters---to assess how closely gem5 matches real hardware behavior~\cite{cebrian2020semi}. However, the critical question of understanding gem5's intrinsic software behavior remains largely under-studied. These behaviors are buried within gem5's modular Python/C++ co-implementation, where complexity arises from event-based scheduling, inter-object communication, and control flow that frequently jumps across different models. \textbf{Conventional profiling tools} also offer limited help. For example, \texttt{gprof}~\cite{gprof1982} produces coarse, flattened call graphs that obscure the layered structure of gem5's abstractions and make it difficult to reason about relationships between simulated components; it also requires code instrumentation, further slowing simulation. Hardware-oriented tools like Intel \texttt{VTune}~\cite{vtune} can collect software call stacks but tend to generate extremely large traces and ultimately suffer from the same limitation: it is hard to explore through gem5's complex call-stack hierarchy at the granularity needed for architectural design-space exploration (DSE).

\textbf{Approach.}
We address this problem with a new holistic profiling toolchain (detailed in~\autoref{sec:mech}) that measures simulation-time breakdowns and reveals how execution time is distributed across gem5's modular components. Our work differs from prior approaches in two key ways:

\begin{itemize}

\item \textit{Lightweight and external.} We use the Linux \texttt{perf\_event} syscall to collect gem5 call-stack traces non-intrusively. Because profiling is performed by a stand-alone helper process rather than by gem5 itself, no instrumentation is added to gem5's execution binary, thereby minimizing overhead and avoiding any perturbation to simulation behavior or performance.

\item \textit{Flexible and hierarchical.} Unlike the flattened call graphs produced by \texttt{gprof} and \texttt{VTune}, our parser preserves the full hierarchical call-tree structure, enabling to isolate specific gem5 execution paths (\eg{}, all functions related to the out-of-order core's \textsf{IEW} stage) while ignoring unrelated functions. It also supports user-defined zoom-in and zoom-out views, allowing both fine-grained analysis of individual modules and holistic examination of cross-components interactions.

\end{itemize}

\textbf{Contributions.}
We apply this toolchain to gem5, focusing on several core models, including \textsf{AtomicSimpleCPU} (\as{}), \textsf{TimingSimpleCPU} (\ts{}), and the \textsf{Out-of-Order CPU} (\ooo{}), together with their interaction with the \ruby{} memory subsystem. We make three primary contributions:

\begin{itemize}

\item \textit{Layer-aware gem5 runtime decomposition.} We provide, to our knowledge, the first comprehensive, layer-aware breakdown of gem5's simulation runtime. Our profiling reveals how gem5 is structured around multiple layers of high-level software abstractions and measures execution time across all major gem5 components (including major core models and their submodules, the \ruby{} memory system, and the \garnet{} interconnection network, \etc{}), revealing cross-component interactions and pinpointing bottleneck functions throughout gem5's software stack.

\item \textit{Demystifying gem5 internals and simulation overheads.}
We systematically explain \textit{where} gem5 spends time across simulated components and \textit{why}. For example, we show that \ts{}---despite its conceptual simplicity and reduced microarchitectural detail---often simulates no faster (and sometimes slower) than a detailed out-of-order core, due to its lockup-cache behavior, in-order pipeline, and busy-wait interaction with \ruby{}. We clarify when and why different gem5 models incur unexpected overheads, providing concrete guidance for selecting appropriate models for architectural DSE.

\item \textit{Correctness debugging beyond statistics.}
We demonstrate that our profiling framework serves as an effective diagnostic tool for gem5 correctness, including the detection of deadlock and livelock in \ruby{}/\slicc{}-based coherence protocols---issues that often do not manifest as explicit simulator errors and are difficult to uncover using gem5 simulation statistics (\texttt{stats.txt}). By identifying characteristic call-stack signatures dominated by repetitive functions and applying per-component runtime-percentage thresholds, our tool can automatically flag potential deadlocks or livelocks and trigger checkpoints, enabling low-overhead root-cause analysis that is difficult to achieve with existing gem5 debugging mechanisms.

\end{itemize}

The rest of the paper is organized as follows:
\autoref{sec:mot} motivates why it is hard to understand gem5 using statistics alone why profiling gem5's call-stack is also challenging.
\autoref{sec:bk} provides background on gem5 components and describes the design of our profiling toolchain and its interaction with gem5.
\autoref{sec:meth} presents our experimental methodology, including simulation configurations and benchmark suites.
\autoref{sec:results} analyzes the profiling results and identifies key performance/correctness bottlenecks.
Finally, \autoref{sec:conc} concludes the paper.

\section{Motivation}
\label{sec:mot}

\subsection{The limits of understanding a simulated system through gem5 statistics}

\begin{figure}[ht]
  \centering
  \includegraphics[width=0.475\textwidth]{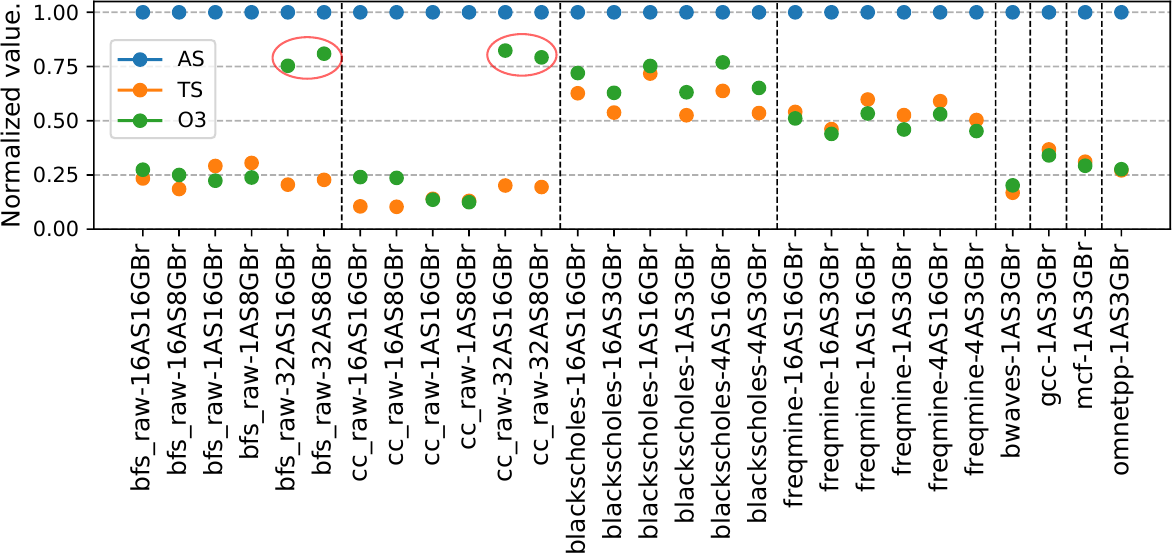}
  \caption{Committed instructions per host-machine-second for the \textsf{AtomicSimple}, \textsf{TimingSimple}, and \textsf{Out-of-Order} gem5 core models across multiple programs, core counts, and \ruby{} memory configurations (details in~\autoref{sec:meth}).}
  \label{fig:commitInsts-model}
\end{figure}

gem5 simulations typically produce a large number of statistics in text-based files (\texttt{stats.txt}). Although these results describe \textit{what} happened---\eg{}, how many instructions have been simulated---they do not explain \textit{how} these numbers are got or \textit{which} simulated components are responsible. For example,~\autoref{fig:commitInsts-model} reports the number of committed instructions per host-machine-second, which indicates how much simulated progress the program has made. In the figure, we normalize all results to AtomicSimpleCPU (\as{}), the simplest core model in gem5. As expected, we observe that TimingSimpleCPU (\ts{}) and the Out-of-Order CPU (\ooo{}) simulate more slowly than \as{}. However, more interestingly, when comparing \ts{} and \ooo{}, we find that \ooo{} is not always slower---even though it models a much more complex core than \ts{}. Actually, in 15 out of 28 cases, \ooo{} simulates faster than \ts{}; in other cases, they achieve close results.

\textbf{Key insight.} These observations raise several critical questions. First, why does \ooo{} sometimes achieve faster simulation progress than \ts{}, even though \ooo{} models many more microarchitectural events---and therefore executes more functions---than \ts{}? Second, in cases where \ts{} and \ooo{} exhibit similar simulation results, does this suggest that \ts{} captures microarchitectural behavior at a level comparable to \ooo{}? If so, can \ts{} be considered a ``good enough'' approximation of \ooo{} for certain studies? 
None of these questions are easy to answer by inspecting gem5's statistics output alone. Instead, our key insight is that, because gem5 simulates a hardware system, its \textit{own} call-stack effectively reveals the internals of the simulated system. Building on this insight, we introduce a \textit{lightweight}, \textit{non-intrusive} toolchain that dynamically samples and analyzes gem5 call-stacks, providing a new way to efficiently and informatively examine gem5's internal behavior.

\textbf{A use case.} Applying our toolchain to the results in~\autoref{fig:commitInsts-model}, we draw two key conclusions that address the above questions. First, although \ts{} simulates fewer microarchitectural events, its in-order pipeline, lockup cache behavior, and lack of instruction-level parallelism (ILP) make each instruction \emph{more expensive} to simulate in time (\autoref{sec:ts}), leading to a runtime comparable to that of a much more detailed \ooo{} core. Consequently, \ts{} is not a ``good enough'' proxy for \ooo{}, even in cases where they achieve a similar number of committed instructions.

Second, simulation results using \ooo{} must be interpreted with care. In four cases in~\autoref{fig:commitInsts-model} (highlighted in red circles), \ooo{}'s higher committed-instruction rate is largely caused by test-and-test-and-set (TTAS) style busy-waiting~\cite{1677499}, where threads repeatedly spin on locks and commit many test loads without making real forward progress (\autoref{sec:ooo}). Consequently, with \ooo{} cores, simulation results should be interpreted together with other system activities---\eg{}, cache-controller actions in~\autoref{fig:O3-L1CC}---to avoid misleading conclusions on performance gains. Our toolchain makes such examinations substantially easier and more flexible than relying solely on gem5's statistics alone.

\subsection{Why is it challenging to profile gem5 call-stacks?}

\begin{figure}[ht]
  \centering
  
  \begin{subfigure}{0.45\textwidth}
    \includegraphics[width=0.976\textwidth]{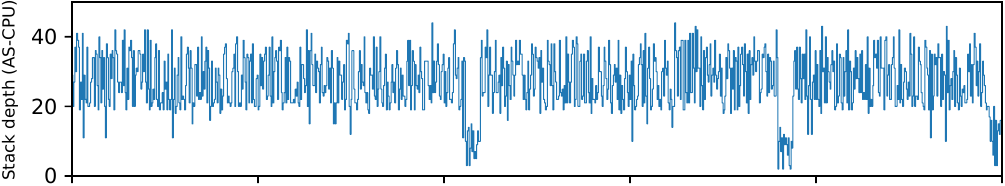}
    \vspace*{0.15 cm} 
  \end{subfigure}

  \begin{subfigure}{0.45\textwidth}
    \includegraphics[width=0.976\textwidth]{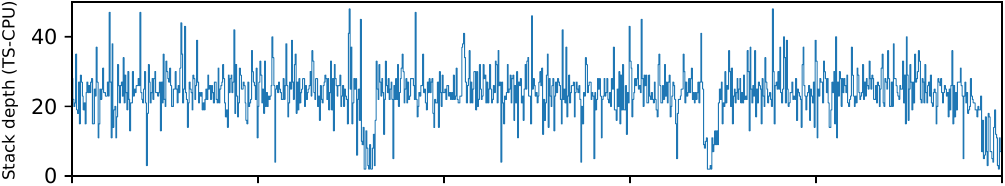}
  \end{subfigure}

  \begin{subfigure}{0.45\textwidth}
    \includegraphics[width=\textwidth]{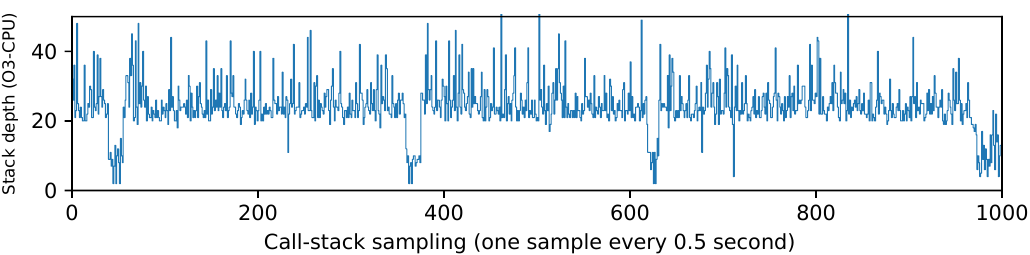}
  \end{subfigure}
  
  \caption{gem5 call-stack depth across CPU models (with \ruby{}).}
  \label{fig:call-stack}
\end{figure}

Interpreting gem5's call-stacks, however, is challenging. The call-stack evolves continuously as gem5 processes events from cores, \ruby{}, and \garnet{}. On one hand, this produces highly fluctuating stack depths (\autoref{fig:call-stack}), with the number of frames frequently swinging between nearly 40 and 0. For example, when an instruction fetch or a data-memory instruction completes in \as{} or \ts{}, or when the \ooo{} transitions between the core \texttt{EventQueue} and the \ruby{} \texttt{EventQueue}, the call-stack becomes shallow because the current call-chain temporarily collapses before gem5 begins processing the next event.

On the other hand, many dominant functions in the stack are merely bookkeeping routines that provide little architectural insight, causing important simulation events to appear only briefly and to be buried deep within the depth. For example, on average, roughly 20 frames in a typical gem5 call-stack originate from embedded Python (\texttt{pybind11}) used to support gem5's Python-based configuration, rather than from any specific microarchitectural component. As a result, gem5's call-chains tend to be deep and noisy, making them difficult to relate to concrete microarchitectural simulations.

Overall, these challenges motivate the specialized tool introduced in this work, which extracts meaningful patterns from raw call-stack traces and makes gem5's behavior more interpretable.

\section{Design and Implementation}
\label{sec:bk}

We next describe how \as{}, \ts{}, and \ooo{} differ in their execution models and how they interact with \ruby{}. We then introduce the design of our profiling toolchain and explain how it works with gem5.

\subsection{AtomicSimpleCPU (\as{})}

\begin{figure}[!h]
  \centering
  \includegraphics[width=\linewidth]{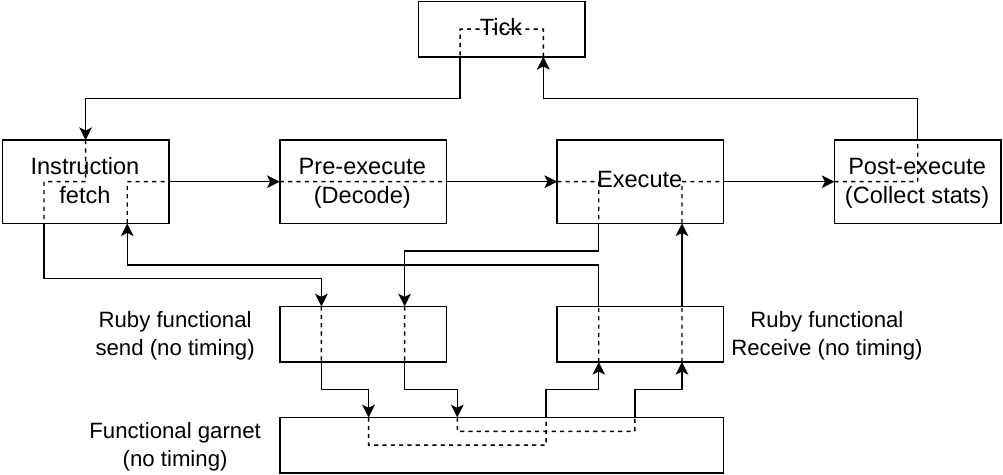}
  \caption{Interaction of \textsf{AtomicSimpleCPU} and \ruby{}, where the execution flows from a core into \ruby{} as a function call (no timing modeled).}
  \label{fig:AS_highlvl}
\end{figure}

\as{} uses a single entry function---the \texttt{tick} function---to drive simulation. On each call, \texttt{tick} \textit{sequentially} advances all core stages, one after the other: \textsf{Instruction fetch}, issuing a memory request to \ruby{} to fetch the instruction from the I-cache; \textsf{Pre-execute}, decoding the instruction; \textsf{Execute}, updating the architectural state (register file) and issuing additional \ruby{} requests for load/store instructions to the D-cache; and \textsf{Post-execute}, committing the instruction and updating internal gem5 statistics. For both instruction and data accesses, TLBs are also invoked for address translation. 

As shown in~\autoref{fig:AS_highlvl}, in \as{}, core stages are implemented as a chain of function calls. Communication between the core and \ruby{} is modeled using function-call as well, bypassing the on-chip interconnect \garnet{}. Because each function runs to completion before returning to its caller, simulated timing is fully abstracted, and instructions execute strictly in program order---after an instruction completes, control returns to the top-level \texttt{tick} function, which initiates the life cycle of the next instruction.

\subsection{TimingSimpleCPU (\ts{})}
\label{sec:ts}

\begin{figure}[!t]
  \centering
  \includegraphics[width=\linewidth]{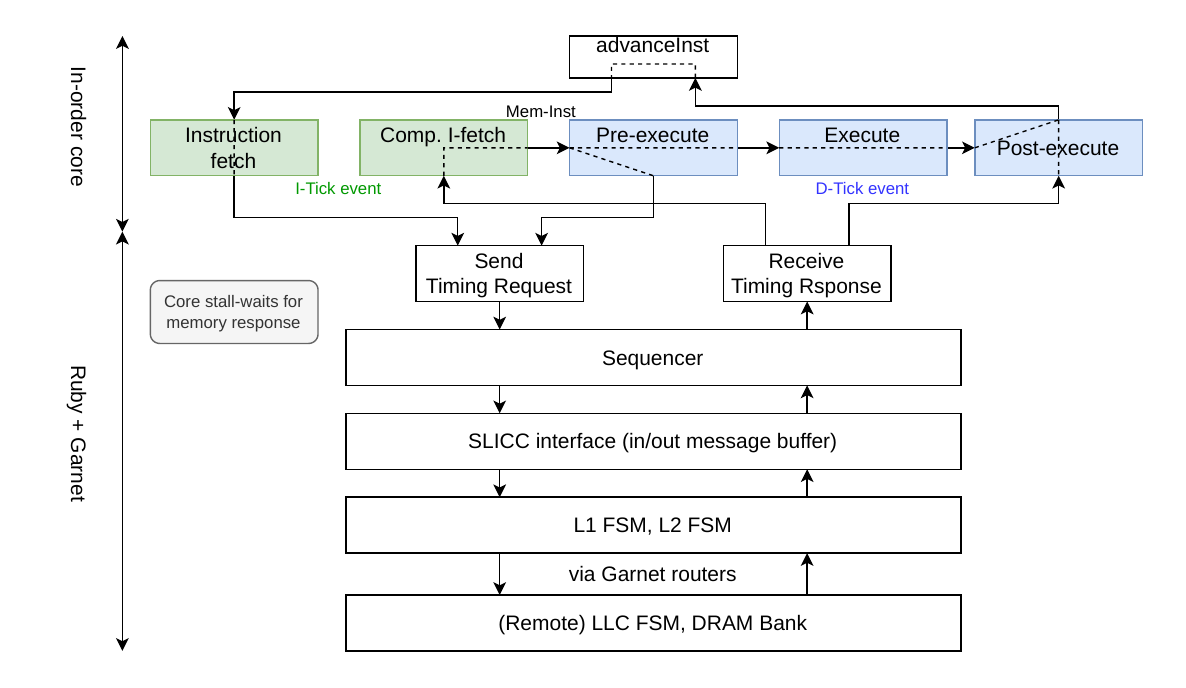}
  \caption{Interaction of \textsf{TimingSimpleCPU} and \ruby{}, where the execution flow between a core and \ruby{} is decoupled by the \ruby{} \texttt{EventQueue}.}
  \label{fig:TS_highlvl}
\end{figure}

\ts{} models \emph{timing} for memory accesses by simulating memory events using \ruby{}'s \texttt{EventQueue} (\autoref{fig:TS_highlvl}). As in \as{}, instructions in \ts{} are executed via function calls and run strictly in order, one at a time. Unlike \as{}, however, each memory access in \ts{} exercises the full \ruby{} memory hierarchy with detailed timing, including private caches, the last-level cache (LLC), and the on-chip interconnect.

Specifically, when a core issues a memory request, \ts{} generates a corresponding \ruby{} request packet and delivers it to the L1 cache controller (L1 CC). After the request incurs the L1 cache latency, it is routed through the on-chip interconnect---either simple crossbar links or \garnet{} routers~\cite{garnet}, both of which model timing latency---to the LLC and then to the memory controller. Once the request is serviced, \ruby{} creates a response packet and sends it back to the \ts{} core. The response travels the path through the memory hierarchy in reverse and ultimately signals completion of the memory access. Because \ts{} cores do not model out-of-order execution or speculation, a core stalls whenever it is waiting for a memory access (for both I-cache and D-cache) to complete, and no other instructions can make progress until \ruby{} returns a response.

The execution flow of \ts{} is shown in~\autoref{fig:TS_highlvl}. The main entry point is the \textsf{advanceInst} function, which initiates an instruction fetch by creating a \ruby{} request (the \texttt{I-Tick} event) and sending it to the I-cache port. When \ruby{} serves this request and returns a response, the I-cache response port invokes \textsf{completeIfetch} as a callback, which in turn calls the \textsf{pre-execute} function to decode the instruction. If the instruction is a non-memory operation, it is executed immediately and the architectural registers are updated. If the instruction is a load or store, \ts{} creates another \ruby{} request (the \texttt{D-Tick} event) to initiate a D-cache access. On completion, a \textsf{post-execute} callback is triggered to complete execution of the memory instruction, and \ts{} moves to \texttt{advanceInst} for the next instruction.

\begin{tcolorbox}[
    colback=black!5,
    colframe=black,
    arc=1mm,
    boxrule=0.2pt,
    left=2pt,
    right=2pt,
    top=2pt,
    bottom=2pt]
\nextsummary The \ts{} model improves memory-timing fidelity relative to \as{}. However, the core remains in-order, busy-waiting, and stalls on every memory access. In addition, simulations execute more slowly because \ts{} generates and processes substantially more events within \ruby{}.
\end{tcolorbox}

\subsection{Out-of-order CPU (\ooo{})}
\label{sec:ooo}

\ooo{} models a superscalar out-of-order core with a deeply pipelined, stage-based microarchitecture. Unlike \as{} and \ts{}, which execute one instruction at a time in program order, \ooo{} supports speculation and dynamic instruction scheduling, exposing both ILP and MLP at the cost of significantly higher complexity and overhead.

\begin{figure}[!t]
  \centering
  \includegraphics[width=\linewidth]{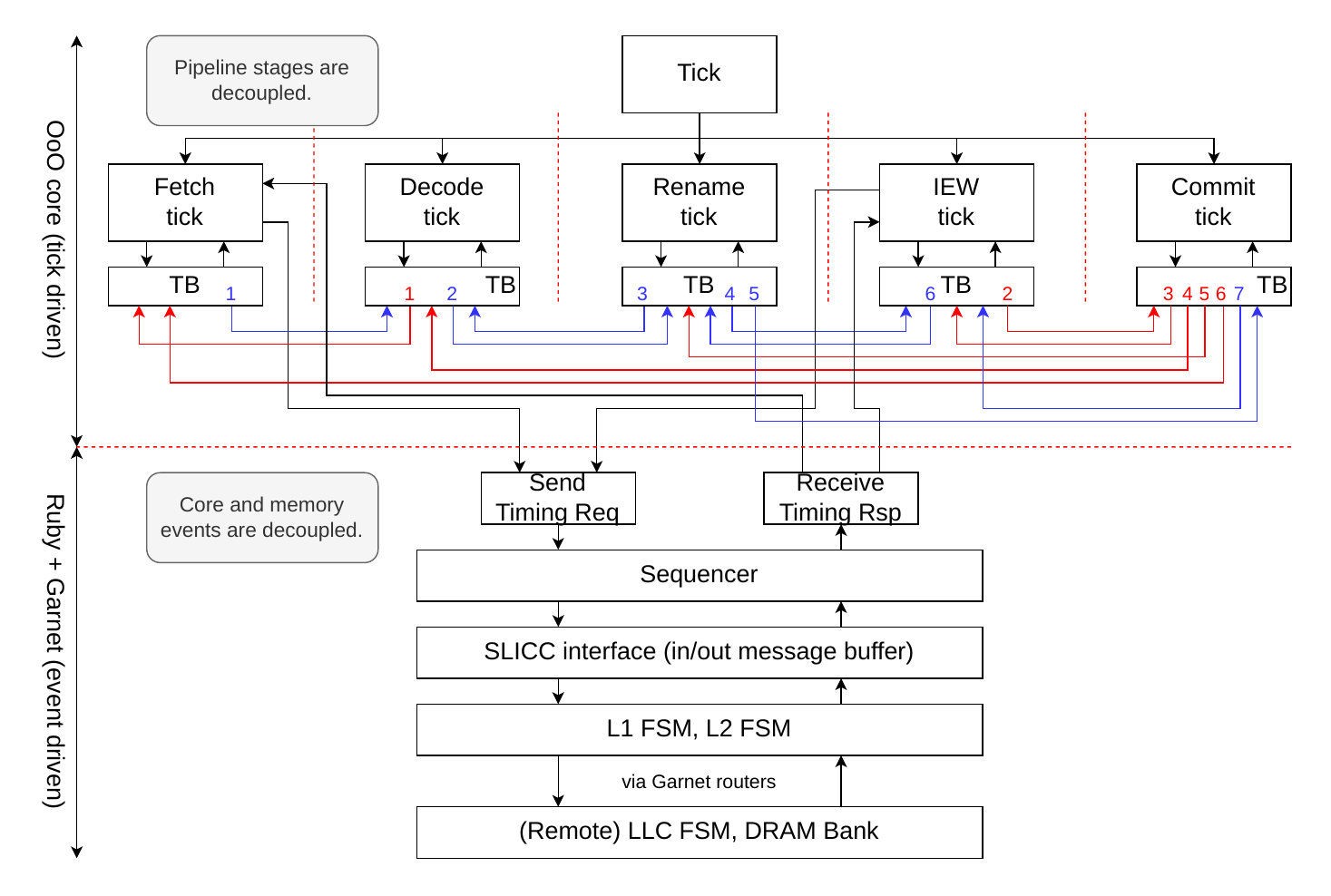}
  \caption{Interaction of the \ooo{} and \ruby{}, where the execution flow between core-stages are decoupled with \texttt{TimeBuffer} to support out-of-order execution.}
  \label{fig:O3_highlvl}
\end{figure}

\begin{table}[!t]
  \centering
  \footnotesize
  \caption{Inter-stage signals used in \ooo{}.}
  \label{tab:squash-signals}
  \begin{tabular}{p{0.14\linewidth}p{0.70\linewidth}}
    \toprule
    \textbf{Category} & \textbf{Description} \\
    \midrule
    \red{Squash-related signals}
    & 1) Branch misprediction detected in \textsf{Decode}. \newline
      2) \textsf{Execute} sends mis-predicted instruction's \texttt{seqNum} and correct PC or violating load/store \texttt{seqNum} (for memory-order violations) to \textsf{Commit}. \newline
      3---6) \textsf{Commit} performs a squash on branch mispredictions, interrupts, or memory-order violations. \\
    \midrule
    \blue{Inter-stage dataflow}
    & 1) Send instructions \textsf{Fetch}~$\rightarrow$ \textsf{Decode}. \newline
      2) Send instructions \textsf{Decode}~$\rightarrow$ \textsf{Rename}. \newline
      3) \textsf{Rename} stall on no free physical registers. \newline
      4) Send instructions \textsf{Rename}~$\rightarrow$ \textsf{IEW}. \newline
      5) Insert instructions into \textsf{ROB}. \newline
      6) \textsf{IEW} updates free entries in \textsf{ROB}/\textsf{LQ}/\textsf{SQ}, and wakes dependents on writeback. \newline
      7) \textsf{Commit} reports number of committed instructions. \\
    \bottomrule
  \end{tabular}
\end{table}

The fundamental difference from \as{}/\ts{} is that, in \ooo{}, both the pipeline stages and the core–memory interaction are \emph{decoupled}. Within the core, gem5 uses cycle-driven simulation: on each cycle, every pipeline stage executes its own \texttt{tick} function, inspects its inputs, and models the corresponding stage behavior. As shown in \autoref{fig:O3_highlvl}, the \ooo{} core comprises five OoO stages: 1) \textsf{Fetch}, retrieves instructions from I-cache and performs an initial decode to construct \emph{dynamic instruction} objects; 2) \textsf{Decode}, completes instruction decoding and resolves unconditional branches; 3) \textsf{Rename}, manages logical-to-physical register mappings, including allocation and reclamation of physical registers; 4) \textsf{IEW} (Issue/Execute/Writeback), schedules ready instructions, dispatches them to functional units, resolves conditional branches, and performs writeback; and 5) \textsf{Commit}, tracks instruction completion, handles exceptions and squashes, and retires instructions from the reorder buffer.

In \ooo{}, pipeline stages are decoupled through inter-stage buffers—specifically, \texttt{TimeBuffer}s (TBs)—with the associated data-flows and squash-signals summarized in~\autoref{tab:squash-signals}. As shown in~\autoref{fig:O3_highlvl}, each stage communicates with others via \texttt{TB}s, enabling stages to progress independently rather than stall while waiting for their neighbors as in \as{} and \ts{}. For example, the red arrow from \textsf{Commit} to \textsf{IEW} indicates that \textsf{Commit} issues a squash signal to \textsf{IEW}. Once the signal is sent to the \texttt{TB}, \textsf{Commit} continues processing subsequent instructions, while \textsf{IEW} performs the required squash when its \texttt{tick} function is next executed.

The core's interaction with \ruby{} is also decoupled. Like \ts{}, \ooo{} uses \ruby{} for timing memory accesses in \textsf{Fetch} and \textsf{IEW}. However, because instructions in \ooo{} execute out-of-order, the \textsf{IEW} stage also sends speculative memory requests to \ruby{} and enforces memory-ordering constraints defined by a system's memory model. For example, under the x86 TSO model in gem5, both loads and stores are executed speculatively, with their ordering constraints checked by the \ooo{}. For example, if 1) a younger load executes and then becomes invalidated, or 2) a younger load executes before an older store to the same address executes, the younger load will be marked for squashing and enqueued into the core's replay buffer.
Such behavior increases simulated program performance (\eg{}, ILP) but incurs the number of gem5 runtime events, as more core pipeline and memory activities will be simulated.

\begin{tcolorbox}[
    colback=black!5,
    colframe=black,
    arc=1mm,
    boxrule=0.2pt,
    left=2pt,
    right=2pt,
    top=2pt,
    bottom=2pt]
\nextsummary The \ooo{} model increases the instruction-level parallelism (ILP) of simulated programs and improves benchmark performance; however, it also generates substantially more events than \ts{} within the out-of-order pipeline and in \ruby{} due to speculative instruction execution.
\end{tcolorbox}

\subsection{Design of the gem5 profiling toolchain}
\label{sec:mech}

\begin{figure}[!t]
  \centering
  \includegraphics[width=\linewidth]{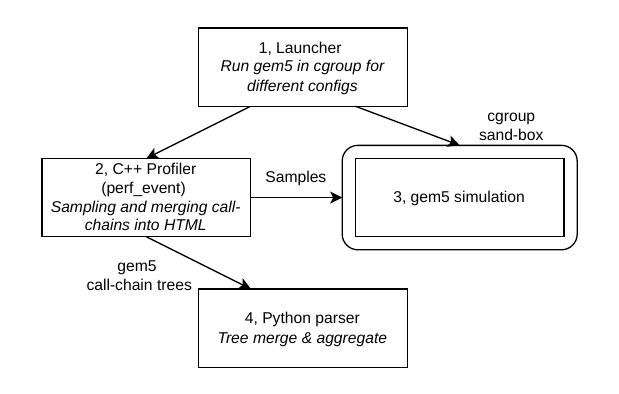}
  \caption{Overview of the profiling and analysis toolchain used for gem5 experiments.}
  \label{fig:toolchain}
\end{figure}

To examine the simulated core microarchitecture and its interactions with \ruby{}, we develop a lightweight profiling toolchain that supports fine-grained, flexible gem5 call-stack profiling. This makes it well suited for both performance debugging and architectural analysis. As shown in \autoref{fig:toolchain}, the toolchain comprises three components: 1) a launcher that starts and manages gem5 runs under different simulation configurations, 2) a profiler that periodically samples gem5 call stacks, and 3) a call-stack analyzer that processes the profiled call-stacks at multiple levels of granularity. We next describe the design of each component.

\textbf{The launcher} first creates a Linux \texttt{cgroup} and then runs a gem5 process within it to constrain resource usage and isolate the gem5 process from other system activities. Inside the \texttt{cgroup}, a gem5 process picks up a benchmark program from \autoref{tab:benchmarks} in full-system simulation, with a specified gem5 configuration shown in~\autoref{tab:core-config-zen5}.

\begin{figure}[!t]
  \centering
  \includegraphics[width=\linewidth]{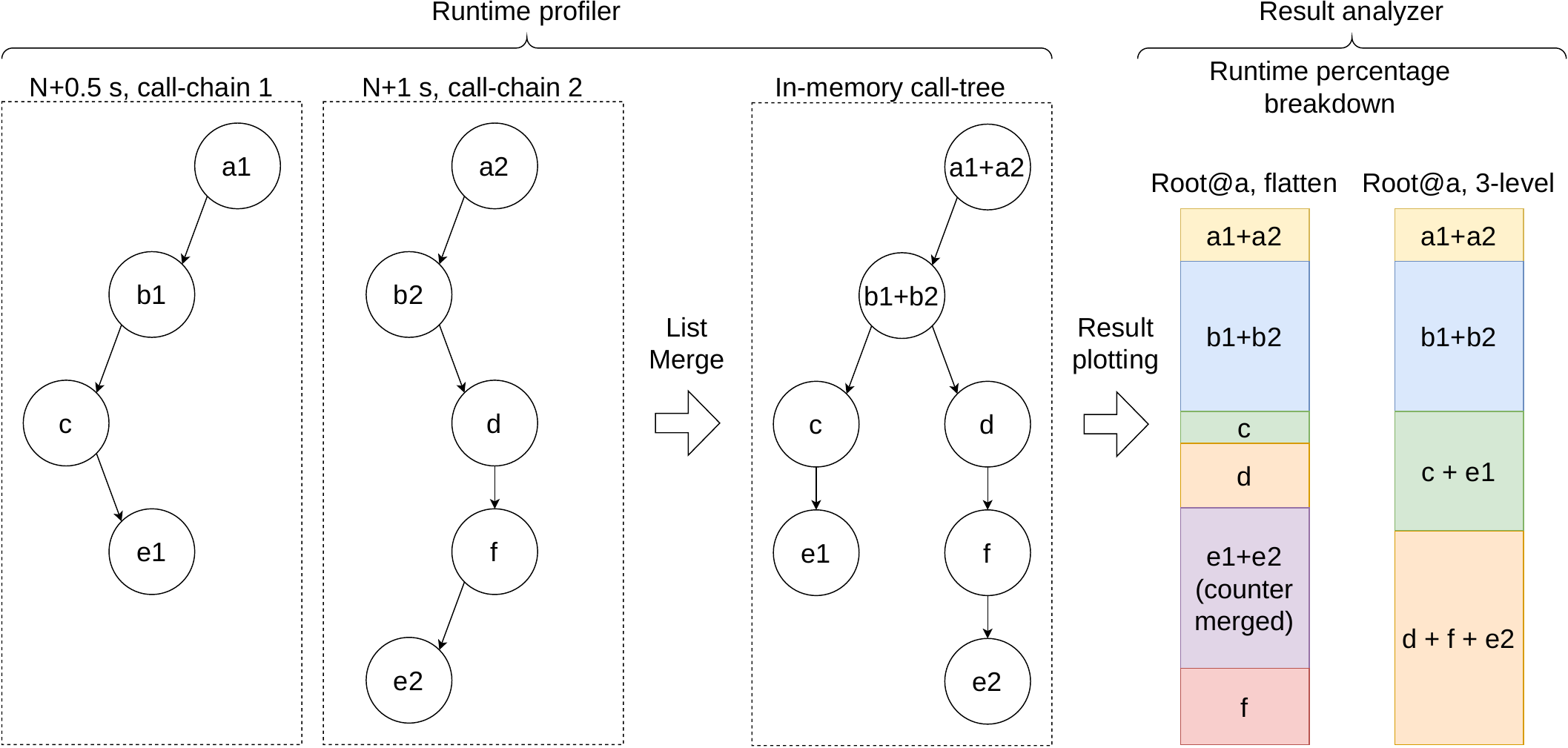}
  \caption{Call-stack merging and flexible view-control in the call-stack analyzer.}
  \label{fig:tree}
\end{figure}

The \textbf{profiler} is sampling-based and built on Linux's \texttt{perf\_event} interface. It periodically samples a running gem5 process and captures its call-chains. Given a gem5 PID, the profiler first locates the corresponding \texttt{perf\_event} cgroup via \texttt{/proc/<pid>/cgroup}, then attaches to that cgroup using \texttt{perf\_event\_open}, thereby profiling all threads and descendant processes within the same cgroup.

To resolve symbols, the profiler dynamically loads function symbols from the gem5 binary and all linked third-party libraries. It scans \texttt{/proc/<pid>/maps}, loads ELF images via \texttt{/proc/<pid>/root}, and constructs address-to-symbol mappings from the \texttt{SHT\_SYMTAB} and \texttt{SHT\_DYNSYM} tables. Function names are demangled using \texttt{abi::\_\_cxa\_demangle}, and kernel frames are resolved via \texttt{/proc/kallsyms}. The main gem5 process in the cgroup exposes a shared ring buffer \texttt{perf\_event\_mmap\_page}, from which the profiler polls and parses records (\texttt{PERF\_RECORD\_SAMPLE}) containing gem5 call chains. We configure the profiler to use the software CPU clock (\texttt{PERF\_TYPE\_SOFTWARE}, \texttt{PERF\_COUNT\_SW\_CPU\_CLOCK}) with a moderate sampling period (0.5\,s by default).

During call-stack sampling, each individual sample is first processed as a linked list and then merged into an in-memory call-tree data structure (as illustrated in~\autoref{fig:tree}). In the call tree, call stacks that share a common prefix are merged and diverge at their first differing branch. For example, \texttt{a1->b1->c->e1} and \texttt{a2->b2->d->f->e2} share the prefix \texttt{a->b}; during merging, the counters for these shared nodes accumulate to \texttt{a1+a2} and \texttt{b1+b2}, respectively. After the shared prefix, the paths split, and the same callee (\eg{} \texttt{e1} and \texttt{e2}) reached from different callers (\texttt{c} and \texttt{f}, respectively) is treated as originating from distinct call sites, with counters maintained separately.

At the end of a simulation, the profiler exports the collected call tree as an interactive HTML/JSON report that enables users to visually explore internal gem5 events. The report presents hierarchical call stacks consistent with~\autoref{fig:tree} and can be interactively expanded or collapsed to quickly identify hot simulation events---which are often difficult to extract from the gem5 statistics file or from traditional flat profiles generated by tools such as \texttt{gprof}.

Finally, the \textbf{call-stack analyzer} processes the HTML/JSON call-stack traces at different granularities, for which the root of interest and the maximum depth can be configured. For example,~\autoref{fig:tree} shows 1) a \emph{flattened} view, in which all nodes are shown and counters for identical functions are merged, and 2) a \emph{3-level} view, in which the tree is truncated to show at most three levels and all deeper nodes are aggregated into their ancestor at the last level (\eg{}, nodes \texttt{c} and \texttt{e1} are merged into \texttt{c}; \texttt{d}, \texttt{f}, and \texttt{e2} are merged into \texttt{d}). This flexibility enables rapid navigation between overview and detail. In addition, the call-stack analyzer automatically generates result figures, some of which are used in the evaluation section.

\section{Methodology}
\label{sec:meth}

\subsection{Simulated architecture}

We apply the proposed toolchain through extensive simulations on gem5 full-system mode (FS), running a diverse set of benchmarks across multiple gem5 configurations. We use the x86\_64 ISA with three core models: \as{}, \ts{}, and \ooo{}. For \ooo{}, we use the state-of-the-art AMD Zen5 microarchitectures (\autoref{tab:core-config-zen5}).
For the memory system, \as{} uses functional \ruby{}, whereas \ts{} and \ooo{} use \ruby{}+\garnet{}. All networks use 128-bit links and 1-cycle routers, together with a directory-based 2-level MOESI coherence protocol and 64\,Byte cacheline.

\begin{table}[!t]
  \centering
  \caption{Modeled \ooo{} microarchitecture (AMD Zen5).}
  \label{tab:core-config-zen5}
  \footnotesize
  \setlength{\tabcolsep}{3pt}
  \begin{tabular}{@{}llll@{}}
    \toprule
    \textbf{Parameter} & \textbf{Value} &
    \textbf{Parameter} & \textbf{Value} \\
    \midrule
    Fetch/Decode & $8 / 8$ &
    L1I/assoc. & $32$\,kB/$8$ \\
    Rename & $8 / 8$ &
    L1D/assoc. & $48$\,kB/$8$\textsuperscript{*} \\
    Issue/Commit & $8 / 8$ &
    L2/assoc. (LLC) & $1$\,MB/$16$ \\
    ROB/IQ entries & $320 / 128$ &
    NoC & Mesh X-Y \\
    LQ/SQ entries & $64 / 64$ &
    Link/Router & 128-bit link/1-cycle router \\
    Phys.\ int/FP reg & $256 / 256$ &
    BP/BTB & 64\,Kbit TAGE\_SC\_L/4096 \\
    \bottomrule
  \end{tabular}
  \raggedright\footnotesize
  \textsuperscript{*}Associativity rounded to the nearest power of 2 to satisfy gem5's Tree-PLRU replacement policy.
\end{table}

\subsection{Benchmarks}

\begin{table}[!t]
  \centering
  \caption{Benchmarks.}
  \label{tab:benchmarks}
  \footnotesize
  \begin{tabular}{lccccc}
    \toprule
    \textbf{Suite} &
    \textbf{Programs} &
    \textbf{Input size} &
    \textbf{Core} &
    \textbf{Memory} \\
    \midrule
    
    GAPBS &
    bfs, cc &
    \texttt{12GB graphs} &
    1/16/32 &
    8/16\,GB \\
    
    PARSEC-3.0 &
    blackscholes &
    \texttt{SimLarge} &
    1/4/16 &
    3/16\,GB \\
    
    PARSEC-3.0 &
    freqmine &
    \texttt{SimLarge} &
    1/4/16 &
    3/16\,GB \\
    
    CPU2017 &
    \uline{b}waves\textsuperscript{*}, \uline{g}cc\textsuperscript{*} &
    \texttt{ref} (large) &
    1 &
    3\,GB \\

    CPU2017 &
    \uline{m}cf\textsuperscript{*}, \uline{o}mnetpp\textsuperscript{*} &
    \texttt{ref} (large) &
    1 &
    3\,GB \\
    
    \bottomrule
  \end{tabular}
  \raggedright\footnotesize
  \textsuperscript{*}Because of space constraints, the result figures use the abbreviations \textsf{b}, \textsf{g}, \textsf{m}, and \textsf{o} to represent these programs.
\end{table}

We use three benchmark suites: GAPBS~\cite{gapbs} for multithreaded graph analytics workloads, PARSEC-3.0~\cite{parsec3} for general-purpose multithreaded applications, and SPEC CPU2017~\cite{spec2017} for CPU-intensive single-threaded workloads. From these suites, we select \texttt{bfs} and \texttt{cc} from GAPBS, \texttt{blackscholes} and \texttt{freqmine} from PARSEC-3.0, and \texttt{bwaves}, \texttt{gcc}, \texttt{mcf}, and \texttt{omnetpp} from SPEC CPU2017. As shown in~\autoref{tab:benchmarks}, each program uses a large input set and is evaluated under multiple gem5 configurations, with varying number of cores as well as the total system memory size.

For GAPBS and PARSEC-3.0 programs, we employ 2$\times$2, 4$\times$4, and 4$\times$8 meshes with 4-, 16-, and 32-cores, respectively. Further, each experiment samples gem5 call-stack after the program reaches its region of interest (ROI). For SPEC CPU2017, we use the SimPoint methodology~\cite{simpoint} to generate checkpoints and resume execution from the SimPoint with the highest weight, representing the dominant phase of a program. We use 5\,Million instructions for warmup and 100\,Million instructions for simulation. Throughout the evaluation, x-axis labels in the result figures encode the core–memory configuration of each run. For example, \texttt{4TS3r} denotes a configuration with 4 \ts{} cores, 3\,GB of memory, and Ruby-enabled memory (\texttt{r}).

\section{Evaluation results}
\label{sec:results}

\subsection{Results for \as{}}

\begin{figure}[!t]
  \centering
  \begin{subfigure}{0.45\textwidth}
    \centering
    \includegraphics[width=\textwidth]{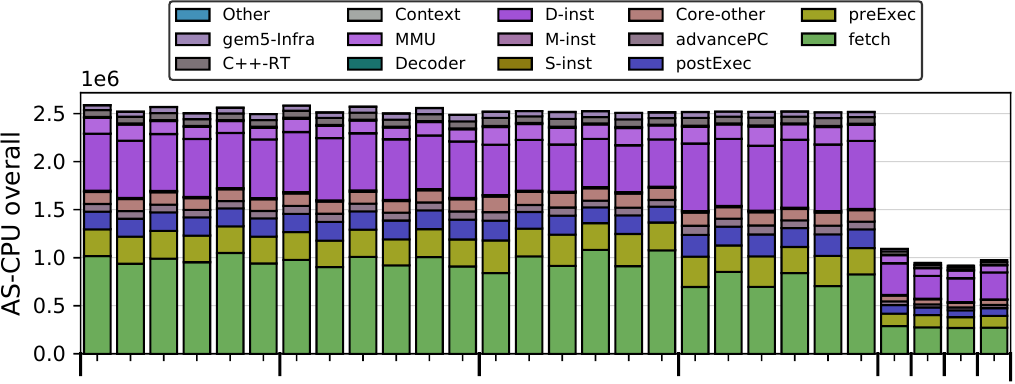}
    \caption{\as{}'s {\tt tick} function runtime breakdown.}
    \label{fig:AS-overall}
    \vspace*{0.25 cm}
  \end{subfigure} 
  \begin{subfigure}{0.45\textwidth}
    \centering
    \includegraphics[width=\textwidth]{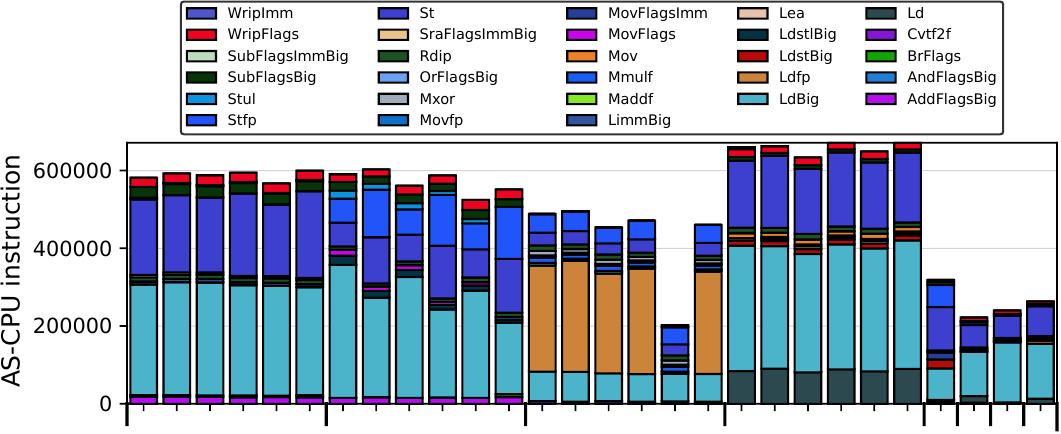}
    \caption{The \texttt{D-inst} stage runtime for \as{}.}
    \label{fig:AS-d-inst}
    \vspace*{0.25 cm}
  \end{subfigure} 
  \begin{subfigure}{0.45\textwidth}
    \centering
    \includegraphics[width=\textwidth]{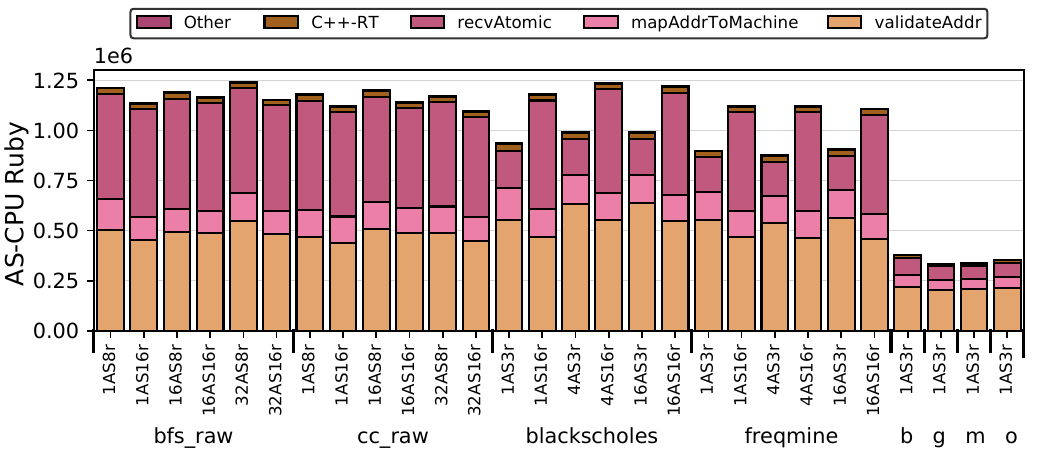}
    \caption{\as{}'s \ruby{} execution runtime breakdown.}
    \label{fig:AS-Ruby-RR}
  \end{subfigure}
  \caption{\as{} runtime breakdown results.}
  \label{fig:AS:ae}
\end{figure}

\autoref{fig:AS-overall} shows the call-stack breakdown of \as{}'s top-level \texttt{tick} function. We observe that, first, compared with GAPBS and PARSEC, SPEC CPU2017 benchmarks generate fewer function calls. This is because all four SPEC workloads complete their simulated instructions (5 Million) within the sampling period, suggesting that \as{} handles single-threaded SPEC applications efficiently and with relatively low simulation cost. Second, across different configurations, gem5 runtime is consistently dominated by \texttt{fetch} and \texttt{D-inst}, where \texttt{fetch} denotes the instruction-fetch stage, and \texttt{D-inst} (dynamic instructions) represents the execution of simulated instructions, including both memory and non-memory instructions. As previously discussed in~\autoref{fig:AS_highlvl}, both stages request accesses to \ruby{} via busy-waiting function calls, which are expensive to simulate. Second, other functions in \as{} contribute limited to overall execution time, including \texttt{S-inst} (static instructions, \ie{}, binary-code translation from a program binary into gem5), \texttt{M-inst} (macro instructions, micro-programs defining complex operations such as \texttt{syscall}), \texttt{preExec} and \texttt{postExec}, \etc{}.

\autoref{fig:AS-d-inst} further illustrates the call-stack breakdown for \texttt{D-inst}. As expected, memory operations dominate {\tt D-inst} runtime across all applications and configurations. Among these, \texttt{LdBig} dominates the majority of programs, while \texttt{blackscholes} from PARSEC-3.0 relies heavily on \texttt{Ldfp} (floating-point loads). \autoref{fig:AS-Ruby-RR} examines the execution-time breakdown of \ruby{} for \as{} requests. It highlights that: \texttt{recvAtomic} (functional access latency in \ruby{}), \texttt{mapAddrToMachine} (address-to-DRAM-controller mapping latency), and \texttt{validateAddr} (cycles for distinguishing DRAM and PIO addresses). In PARSEC-3.0, \texttt{recvAtomic} time increases significantly when scaling memory from 3\,GB to 16\,GB, due to the added overhead of accessing a larger memory space managed by \ruby{}. In contrast, GAPBS sees minimal change from 8\,GB to 16\,GB, since memory beyond 3\,GB is managed within a single contiguous range in gem5.

\begin{tcolorbox}[
    colback=black!5,
    colframe=black,
    arc=1mm,
    boxrule=0.2pt,
    left=2pt,
    right=2pt,
    top=2pt,
    bottom=2pt]
\nextobservation In \as{}, the gem5 call-stack breakdown is largely consistent across different applications and therefore fails to reflect program-specific characteristics, as \as{} is overly abstract. Moreover, the runtime overhead of \as{} is largely dominated by functional accesses in \ruby{}, which increases noticeably when memory is composed of non-contiguous segments.
\end{tcolorbox}


\subsection{Results for \ts{}}

\begin{figure}[!t]
  \centering
  \begin{subfigure}{0.45\textwidth}
    \centering
    \includegraphics[width=\textwidth]{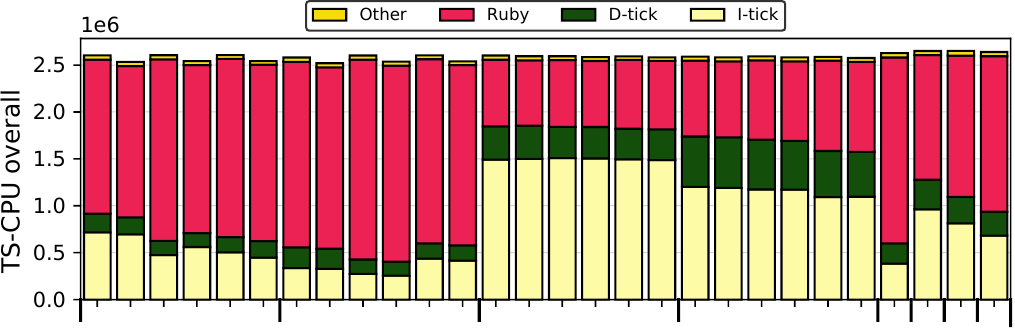}
    \caption{\ts{}'s \texttt{advanceInst} function runtime breakdown.}
    \label{fig:TS-overall}
    \vspace*{0.25 cm}
  \end{subfigure}
  \begin{subfigure}{0.45\textwidth}
    \centering
    \includegraphics[width=\textwidth]{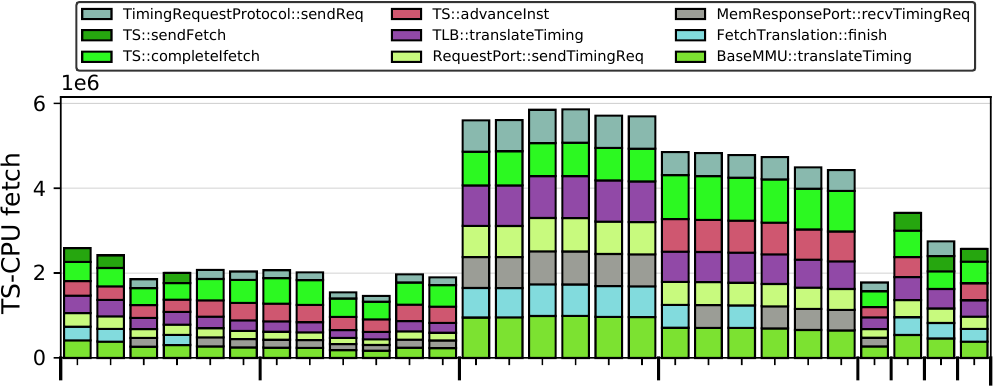}
    \caption{\ts{}'s \texttt{fetch} function runtime breakdown.}
    \label{fig:TS-fetch}
    \vspace*{0.25 cm}
  \end{subfigure}
  \begin{subfigure}{0.45\textwidth}
    \centering
    \includegraphics[width=\textwidth]{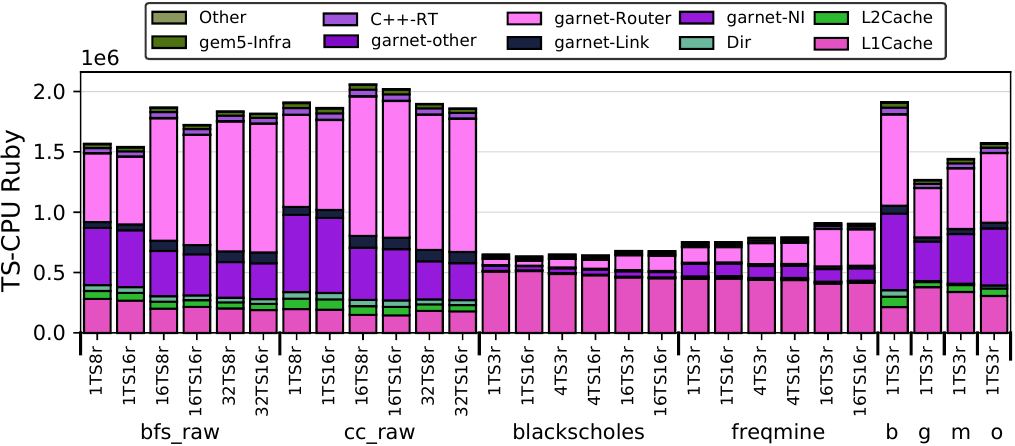}
    \caption{\ts{}'s \ruby{} execution runtime breakdown.}
    \label{fig:TS-ruby}
  \end{subfigure}
  \caption{\ts{} runtime breakdown results.}  
\end{figure}

The execution of \ts{}'s top-level \texttt{advanceInst} function comprises three parts: \texttt{I-tick} (instruction fetch, decoding, and non-memory instruction execution), \texttt{D-tick} (load/store execution), and \ruby{} (memory latency for instruction fetch and load/store), which are discussed in~\autoref{fig:TS_highlvl}, with results shown in~\autoref{fig:TS-overall}.

Compared with \as{}, we observe that \emph{none} of the SPEC CPU2017 benchmarks complete under \ts{} during the sampling period. This is because the detailed \ruby{} model in \ts{} generates substantially more simulation events, significantly slowing simulation progress. Moreover, unlike \as{}---which fails to differentiate program characteristics---the runtime breakdown under \ts{} varies markedly across workloads and begins to reflect differences in applications, as \ts{} uses a finer-grained timing model of \ruby{}. 

For example, for GAPBS, gem5 spends a larger fraction of time in \ruby{}, indicating that these benchmarks are memory-bound. This is because graph workloads exhibit irregular memory accesses and higher LLC miss ratio, leading to substantial time spent waiting for \ruby{} to service memory requests. In contrast, for PARSEC, gem5 spends more time in \texttt{I-tick}, indicating that they are compute-bound, as \texttt{I-tick} includes time spent executing non-memory instructions. The SPEC CPU2017 benchmarks are in general memory-bound, but exhibit varying degrees of memory intensity.

To further investigate the observed trends, our toolchain supports zooming into each part in~\autoref{fig:TS-overall}. For example, \autoref{fig:TS-fetch} presents a zoomed-in view of \texttt{fetch}. We observe that PARSEC workloads generate more \texttt{fetch} events compared to others, with substantial time spent in \texttt{sendReq}/\texttt{sendTimingReq} to access the I-cache, \texttt{completeIfetch} to executing non-memory instructions, and \texttt{BaseMMU/TLB::translateTiming} to access I-TLB. Memory-related instructions are forwarded to \texttt{D-tick} via \texttt{RequestPort::sendTimingReq} and then issued to \ruby{}. Meanwhile, \autoref{fig:TS-ruby} shows a zoomed-in view of \ruby{}, where GAPBS spends significantly more time in \texttt{garnet-Router} and \texttt{garnet-NI} (network interface), indicating heavy stress on the cache hierarchy and on-chip network, whereas PARSEC devotes only a limited fraction of time to \ruby{}. The SPEC CPU2017 benchmarks fall between these two extremes, exhibiting a more balanced mix of compute-bound and memory-bound behavior.

\begin{figure}[!t]
  \centering
  \begin{subfigure}{0.45\textwidth}
    \centering
    \includegraphics[width=\linewidth]{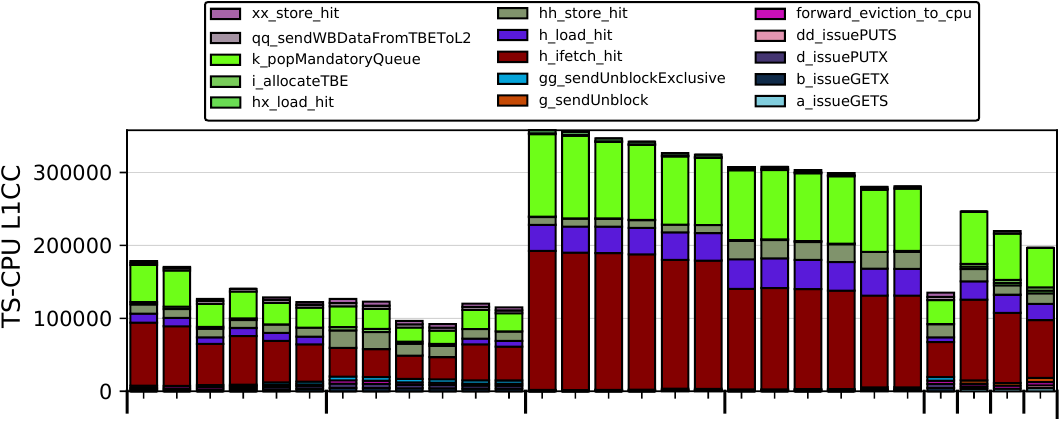}
    \caption{\ts{}'s L1 cache controller actions.}
    \label{fig:TS-L1CC}
    \vspace*{0.25 cm}
  \end{subfigure}
  \begin{subfigure}{0.45\textwidth}
    \centering
    \includegraphics[width=\linewidth]{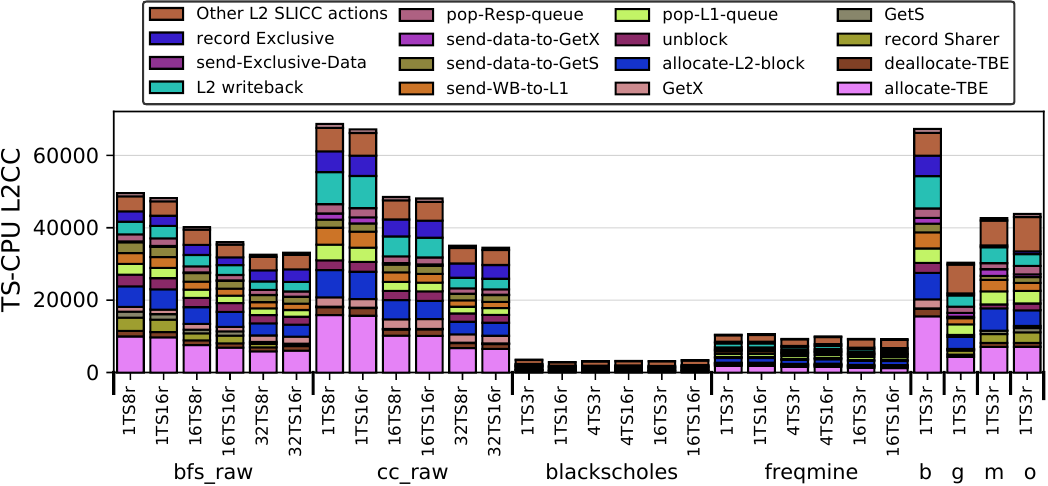}
    \caption{\ts{}'s L2 cache controller actions.}
    \label{fig:TS-L2CC}
    \vspace*{0.25 cm}
  \end{subfigure}
  \caption{\ts{} runtime breakdowns for L1/L2 cache.}  
\end{figure}

Going one level deeper, \autoref{fig:TS-L1CC} and \autoref{fig:TS-L2CC} show the \slicc{}~\cite{slicc, 1206999} actions\footnote{\slicc{} is a domain specific language for specifying cache coherence protocols in gem5.} of the L1 and L2/LLC cache controllers (namely L1CC and L2CC). As expected, PARSEC workloads keep most of their activity within the L1 caches, with a large fraction attributed to \texttt{ifetch-hit}. In contrast, GAPBS generates substantially more traffic to the L2 cache, reflecting higher LLC pressure caused by the irregular memory accesses during graph traversal.

\begin{tcolorbox}[
    colback=black!5,
    colframe=black,
    arc=1mm,
    boxrule=0.2pt,
    left=2pt,
    right=2pt,
    top=2pt,
    bottom=2pt]
\nextobservation Compared to \as{}, the detailed timing model in \ts{} \ruby{} allows gem5 to distinguish between compute-bound and memory-bound workloads.
\end{tcolorbox}

\begin{tcolorbox}[
    colback=black!5,
    colframe=black,
    arc=1mm,
    boxrule=0.2pt,
    left=2pt,
    right=2pt,
    top=2pt,
    bottom=2pt]
\nextobservation \ts{} does not model ILP or MLP: instructions commit in order and busy-wait on memory accesses. As a result, the cache behaves like a \textit{lockup} cache, even though it is modeled as lockup-free in \slicc{}.
\end{tcolorbox}

\subsection{Results for \ooo{}}

\begin{figure}[!t]
  \centering
  \begin{subfigure}{0.45\textwidth}
    \centering
    \includegraphics[width=\linewidth]{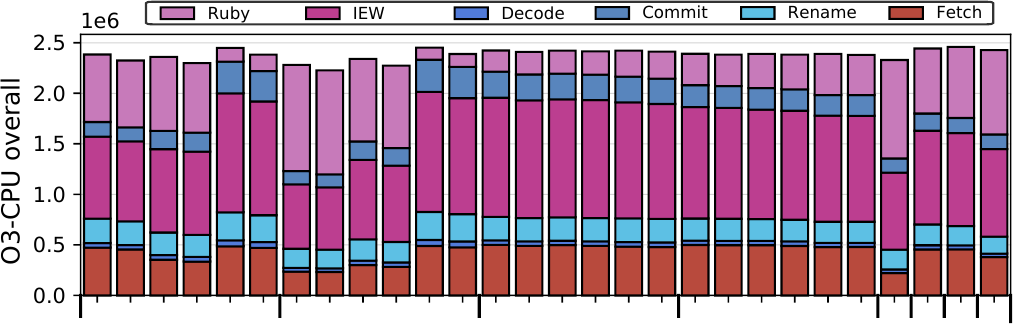}
    \caption{\ooo{}'s \texttt{tick} function runtime breakdown.}
    \label{fig:O3-overall}
    \vspace*{0.25 cm}
  \end{subfigure}
  \begin{subfigure}{0.45\textwidth}
    \centering
    \includegraphics[width=\linewidth]{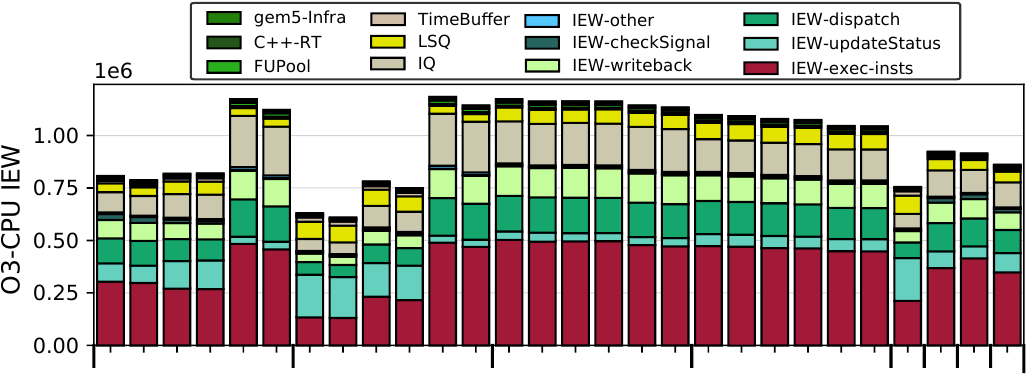}
    \caption{\ooo{}'s \texttt{IEW} stage runtime breakdown.}
    \label{fig:O3-IEW-OA}
    \vspace*{0.25 cm}
  \end{subfigure}
  \begin{subfigure}{0.45\textwidth}
    \centering
    \includegraphics[width=\linewidth]{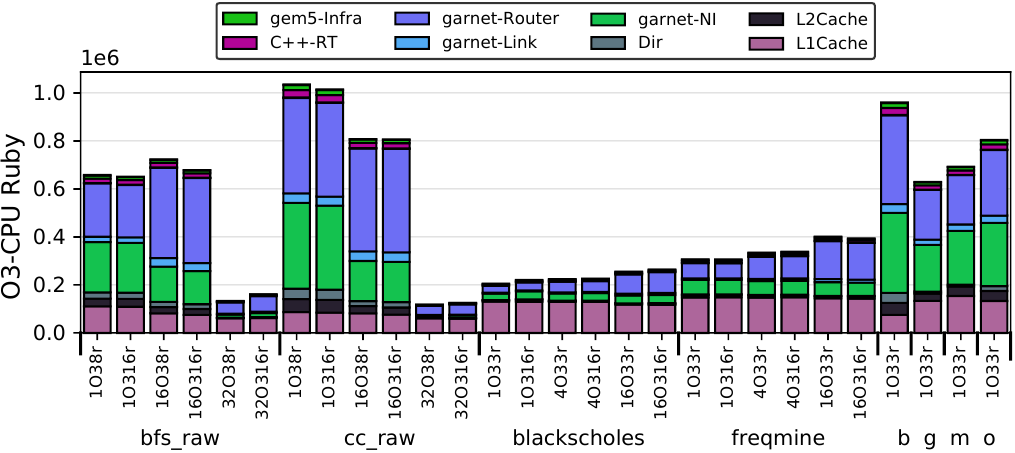}
    \caption{\ooo{}'s \ruby{} execution runtime breakdown.}
    \label{fig:O3-ruby}
  \end{subfigure}
  \caption{\ooo{} runtime breakdown results.}  
\end{figure}

The overall execution time breakdown of \ooo{} is shown in \autoref{fig:O3-overall}. In contrast to \ts{}, \ooo{} models a detailed out-of-order microarchitecture, resulting in a substantial fraction of gem5 runtime spent in the core pipeline stages rather than in \ruby{}. Among these stages, the \texttt{IEW} stage as shown in \autoref{fig:O3-IEW-OA} accounts for the largest share of execution time, reflecting the high modeling cost of instruction scheduling, execution, and writeback in an OoO pipeline.

Another key difference between \ts{} and \ooo{} is how workloads scale with core count. GAPBS and PARSEC exhibit markedly different scaling under \ooo{}. For PARSEC, gem5 runtime breakdown remains stable across the 1-, 4-, and 16-core configurations, indicating limited sensitivity to increased core numbers. In contrast, GAPBS shows substantial divergence in simulation time as core count increases from 16 to 32. In 32-core, \ruby{} (detailed in~\autoref{fig:O3-ruby}) contributes only a negligible fraction of total simulation time, which is effectively the inverse under \ts{}, where \ruby{} dominates gem5 execution for 32-core GAPBS.

Our toolchain enables precise identification of the root causes behind such performance differences, which are difficult to uncover using gem5 statistics alone. In the statistics output, we find that both the instruction and data caches report high hit ratios in this scenario, which can be misleading: it appears that the 32-core configuration benefits from improved cache locality, as if the benchmark partitions a large input into smaller chunks and each core attains higher L1 efficiency.

\begin{figure}[!t]
  \centering
  \begin{subfigure}{0.45\textwidth}
    \centering
    \includegraphics[width=\linewidth]{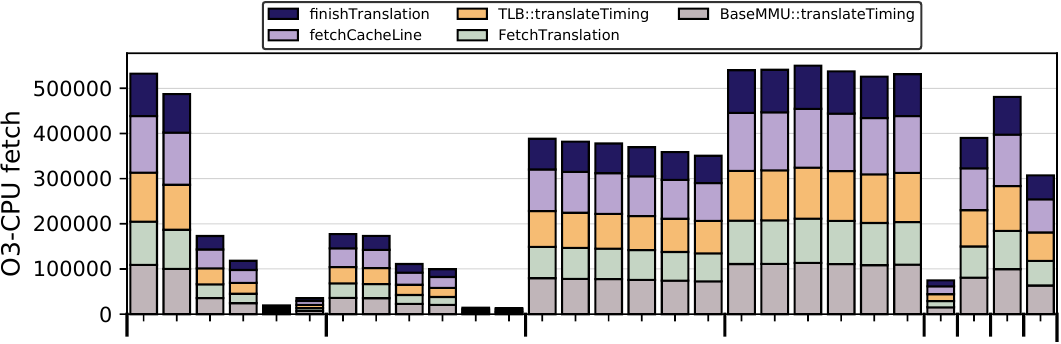}
    \caption{\ooo{}'s \texttt{fetch} function runtime breakdown.}
    \label{fig:O3-fetch}
    \vspace*{0.25 cm}
  \end{subfigure}
  \begin{subfigure}{0.45\textwidth}
    \centering
    \includegraphics[width=\linewidth]{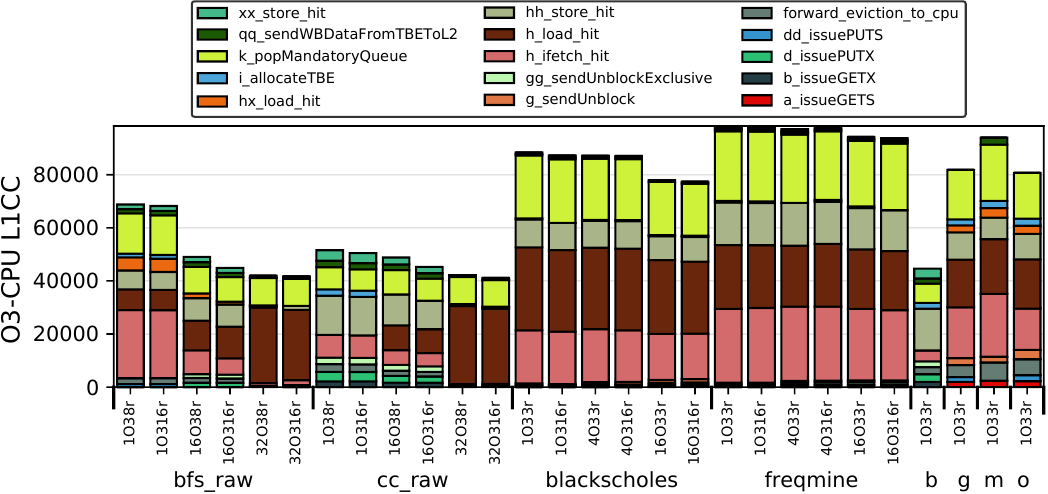}
    \caption{\ooo{}'s L1 cache controller actions.}
    \label{fig:O3-L1CC}
  \end{subfigure}
  \caption{\ooo{} \texttt{fetch} and L1 controller runtime.}
  \label{fig:O3:ae}
\end{figure}

However, the breakdowns produced by our tool---\autoref{fig:O3-fetch} for the \texttt{fetch} stage and \autoref{fig:O3-L1CC} for the L1 cache controller---reveal a different story. In the 32-core GAPBS runs, L1 cache controller runtime is dominated by \textit{data load} events (\texttt{h\_load\_hit}), while \textit{instruction fetch} events (\texttt{h\_ifetch\_hit}) account for only a small fraction of L1 activity. This implies that the core repeatedly executes the same load operations without fetching new instructions from the L1 I-cache, leading to repeated reloads of the same memory address. This is a signature behavior of the \texttt{test-and-test-and-set} (TTAS)-style self-spinning during busy-waiting on a shared lock variable~\cite{1677499}.

By contrast, the PARSEC benchmarks exhibit a far more stable execution profile. Even as core counts increase, \autoref{fig:O3-fetch} and \autoref{fig:O3-L1CC} show that gem5's runtime breakdown remains largely consistent compared to lower-core settings. This stability suggests that \textsf{blackscholes} and \textsf{freqmine} continue to scale effectively with additional cores, spending limited time in self-spinning due to limited number of concurrent data sharers.

\begin{tcolorbox}[
    colback=black!5,
    colframe=black,
    arc=1mm,
    boxrule=0.2pt,
    left=2pt,
    right=2pt,
    top=2pt,
    bottom=2pt]
\nextobservation Compared to \ts{}, \ooo{} simulates the full OoO pipeline, enabling ILP and MLP, and is able to distinguish the same application running across different core counts and memory configurations.
\end{tcolorbox}

\subsection{Detecting coherence protocol deadlock and livelock}

Two problems are particularly important for architectural simulation: cache coherence protocol \textit{deadlock} and \textit{livelock}. In both cases, gem5 makes \textit{no} forward progress and emits \textit{no} explicit failure signal, leaving the simulation either terminated abruptly or running indefinitely. Moreover, existing debugging mechanisms provide limited help in diagnosing protocol dead/livelocks. For example, checkpoint-based diagnosis is ineffective because the onset of a dead/livelock is often unknown. Likewise, running gem5 in debugging mode to monitor repetitive coherence actions provides limited help, as it incurs prohibitive overhead. Consequently, identifying dead or livelock requires monitoring a large span of simulation before the dead or livelock actually happens.

\begin{figure}[!t]
  \centering
  \includegraphics[width=0.45\textwidth]{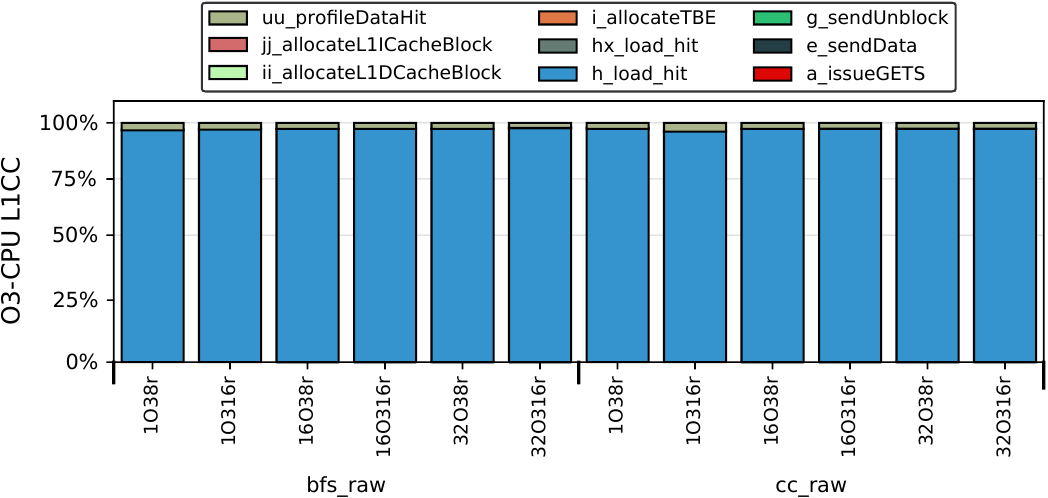}
  \caption{\ooo{} L1 runtime under deadlock.}
  \label{fig:dlll-o3}
\end{figure}

To address this problem, our tool provides a new, low-overhead mechanism for identifying coherence dead/livelock by monitoring the coherence controller runtime breakdown of gem5. The \textbf{key insight} is that when deadlock or livelock occurs, gem5 repeatedly executes the same protocol actions, causing the runtime breakdown dominated by a certain set of functions. Based on this insight, we impose threshold on each runtime proportion for an action (\eg{} 90\%), and when an action exceeds its threshold, the profiler checkpoints the simulation and emits a warning. By using such event-triggered mechanism, we eliminate the high overhead associated with traditional methods for dead/livelock detection.

We demonstrate this mechanism using GAPBS by modifying gem5's \slicc{} code to deliberately inject a protocol-level deadlock: a load request that enters the L1 input (mandatory) queue is never serviced and is instead continuously recycled. As shown in~\autoref{fig:dlll-o3}, this change causes the L1 cache to be overwhelmingly dominated by \texttt{load\_hit}, which happens consistently across different O3 configurations. When the profiler detects such a runtime threshold violation---where a single activity accounts for more than 90\% of execution time---it issues a warning, takes a checkpoint, and reports the event to the user.

\begin{tcolorbox}[
    colback=black!5,
    colframe=black,
    arc=1mm,
    boxrule=0.2pt,
    left=2pt,
    right=2pt,
    top=2pt,
    bottom=2pt]
\nextobservation The proposed toolchain facilitates the investigation of simulation \textit{correctness} issues, including \ruby{} \slicc{} deadlocks and livelocks, and incurs lower overhead than conventional approaches such as debugger-based monitoring.
\end{tcolorbox}

\textbf{Comparison to modifying gem5 internals for deadlock/livelock detection.}
Our approach adopts a non-intrusive profiling methodology, which introduces several trade-offs compared to directly modifying gem5 internals or simulator statistics for deadlock/livelock detection. 

Direct instrumentation of gem5 (\eg{}, adding custom statistics or explicit dead/live-lock detection logic) can provide more precise and robust detection, since developers can directly observe architectural states, such as queues, protocol transitions, or resource dependencies, to examine a deadlock/livelock. However, such approaches typically require prior knowledge of what components in gem5 to instrument.

In contrast, our method infers abnormal simulator behavior by observing distortions in gem5's call-stack profile using external sampling via \texttt{perf\_event}. It is completely non-intrusive and requires no modification to gem5 source code, lowering the barrier for adoption, especially for complex experimental gem5 branches. Moreover, the profiler provides a holistic view into gem5 execution without requiring developers to predefine which components to monitor.

In fact, our approach can serve as a monitoring tool to help pinpoint potential deadlock causes, after which developers can introduce targeted instrumentation inside gem5 for more direct monitoring. In this way, the profiler can guide deeper instrumentation by identifying suspicious hotspots or abnormal execution patterns, helping developers determine where more detailed statistics or internal modifications are needed.

\subsection{Discussions}

\textbf{Limitations.} While the proposed framework provides a lightweight method for analyzing gem5 runtime behavior, it requires some consideration during use. First, the sampling interval must balance detail and runtime overhead. In our implementation, a sampling period of 0.5\,s provides sufficient visibility into gem5 behavior while maintaining negligible overhead during long-running simulations. Nevertheless, very short-lived events may not always be captured. Second, the profiler observes call-stacks of the gem5 process rather than the architectural state of the simulated program. Consequently, the observed execution patterns reflect the simulator implementation rather than the simulated workload directly. Overall, the proposed profiler is not intended to replace internal correctness checks or specialized debugging infrastructure of gem5. Instead, it should be viewed as a complementary tool that helps developers rapidly identify suspicious behaviors and performance bottlenecks in gem5.

\textbf{Identifying gem5 optimization opportunities.}
Beyond debugging, the proposed profiler can also reveal performance optimization opportunities of gem5. For example, we observe that gem5 O3CPU simulations spends a significant portion of runtime creating dynamic instruction objects. In particular, functions related to dynamic instruction construction (\eg{}, \texttt{Fetch::buildInst}) frequently appear among the most heavily executed routines in the call-stack profile. These functions allocate \texttt{DynInst} objects using C++ dynamic memory allocation for each simulated instruction.

This observation suggests a potential optimization opportunity: introducing a \texttt{DynInst} memoization pool. Instead of repeatedly allocating and deallocating dynamic instruction objects, gem5 could reuse a bounded pool of preallocated objects up to the number permitted by the reorder buffer (ROB). Such reuse mechanisms could reduce memory allocation overhead and improve overall simulator runtime efficiency.

\section{Conclusion}
\label{sec:conc}

We demonstrate that call-stack profiling in gem5 provides a flexible, complementary view of simulated architectures that conventional gem5 statistics alone fall short. Using an external sampling-based profiler built on Linux's \texttt{perf\_event} interface, we propose a lightweight and external toolchain that characterizes how execution time is distributed across gem5's core models and the \ruby{} memory system. This toolchain enables fine-grained analysis of runtime differences among core models, helps uncover simulated programs' performance caveats, and supports correctness debugging for gem5, including detection of protocol deadlocks and livelocks.

\section*{Acknowledgments}
We acknowledge the use of computing resources provided by the National Academic Infrastructure for Supercomputing in Sweden (NAISS) under project numbers NAISS 2023/22-1215, NAISS 2024/5-314, and NAISS 2024/6-189. This work is supported by the Swedish Research Council (VR) starting grant number 2025-04436.

\bibliographystyle{IEEEtran}
\bibliography{main}

%
%
%
%
%


\clearpage
\appendix
\section{Artifact Appendix}

\renewcommand*\DTstyle{\small\sffamily}
\subsection{Abstract}

This artifact provides the complete experimental infrastructure used in the paper
\textit{Understanding Simulated Architecture via gem5 Call-Stack Profiling}.
It includes the software components, scripts, and workflows necessary
to reproduce the experimental results. The artifact contains:

\begin{itemize}
\item A modified gem5 source tree (version 23) that contains a non-intrusive call-stack profiler (\texttt{src/ext/prof});
\item Configuration files used to run the gem5 simulations;
\item Scripts for launching experiments using a Python Celery-based task queue system;
\item Parsing scripts for processing gem5 simulation outputs;
\item Plotting scripts that reproduce the figures reported in the paper.
\end{itemize}

Using the provided infrastructure, one can reproduce the main experimental
results shown in~\autoref{fig:AS:ae}--\autoref{fig:O3:ae} of the paper.

\subsection{Artifact check-list (meta-information)}

{\small
\begin{itemize}
  \item {\bf Algorithm: } Call-stack profiling for gem5 and analysis scripts
  \item {\bf Program: } gem5 architectural simulator
  \item {\bf Compilation: } scons build system
  \item {\bf Transformations: } -
  \item {\bf Binary: } gem5.debug, gem5.opt
  \item {\bf Model: } AtomicSimpleCPU, TimingSimpleCPU, O3CPU, Ruby
  \item {\bf Data set: } gem5 checkpoints for: GAPBS benchmarks (\textsf{bfs}, \textsf{cc}), PARSEC-3.0 (\textsf{blackscholes}, \textsf{freqmine}), SPEC CPU2017 (\textsf{bwaves}, \textsf{gcc}, \textsf{mcf}, \textsf{omnetpp})
  \item {\bf Run-time environment: } Linux server environment
  \item {\bf Hardware: } Intel Core i7-12900K, 128\,GB RAM (the paper's platform)
  \item {\bf Run-time state: } full-system gem5 simulations
  \item {\bf Execution: } Celery-based parallel task launcher
  \item {\bf Metrics: } gem5 call-stack profiling statistics, gem5 runtime breakdown
  \item {\bf Output: } gem5 call-stack samples and plots corresponding to~\autoref{fig:AS:ae}--\autoref{fig:O3:ae} (and more, see~\autoref{ae:g})
  \item {\bf Experiments: } 84 gem5 simulation runs across CPU models and benchmark suites
  \item {\bf How much disk space required (approximately)?: } $\sim$500 GB, including gem5 checkpoints, generated data, gem5 disks and Linux kernel binaries
  \item {\bf How much time is needed to prepare workflow (approximately)?: } 10--20 minutes
  \item {\bf How much time is needed to complete experiments (approximately)?: } $\sim$21 hours using 4 parallel workers
  \item {\bf Publicly available?: } Yes
  \item {\bf Code licenses (if publicly available)?: } gem5 open-source license
  \item {\bf Data licenses (if publicly available)?: } research use
  \item {\bf Workflow automation framework used?: } Celery
  \item {\bf Archived (provide DOI)?: } Yes
\end{itemize}
}

\subsection{Description}

\subsubsection{How to access}

The artifact can be downloaded via: 
\begin{itemize}
    \item \url{https://doi.org/10.5281/zenodo.19126063}, or
    \item \url{https://zenodo.org/records/19126063}
\end{itemize}

\subsubsection{Hardware dependencies}

The experiments require a Linux server capable of using \texttt{perf\_event} (this probably requires root privileges, depending on the system configuration) and gem5 full-system simulations. The system used in our evaluation has the following configuration: i) x86 processor (Intel Core i7-12900K), ii) 128\,GB RAM.

The workflow launches multiple gem5 simulations in parallel using Celery workers.
In our experiments we used 4 workers to avoid memory exhaustion during
full-system simulations.

\subsubsection{Software dependencies}

The artifact depends on the following software: i) Linux operating system, ii) Python and Celery task framework, iii) scons build system, iv) gem5 simulator. System monitoring tools such as \texttt{htop} or \texttt{btop} are recommended for observing resource usage during experiments.

\subsubsection{Datasets}

Due to the large sizes (about 350\,GB) of the gem5 checkpoints as well as the disk images and kernel binaries required for the simulations, these files cannot be hosted on Zenodo. \uline{Therefore, users are responsible for preparing them locally.} Checkpoints can be generated by following the tutorials provided by gem5, for example the PARSEC tutorial\footnote{\url{https://www.gem5.org/documentation/gem5art/tutorials/parsec-tutorial}}. Once prepared, the disk and kernel files can be organized using the following directory structure:

\dirtree{%
.1 ckpt/ \DTcomment{simulation checkpoints}.
.2 \red{gapbs-big}/.
.3 program-name.
.4 8GB/16GB.
.5 1p-r1/16p-r1/32p-r1.
.2 \orange{parsec}/.
.3 program-name.
.4 3GB/16GB.
.5 1p-r1/4p-r1/16p-r1.
.2 \blue{spec2017}/.
.3 program-name.
.4 3GB.
.5 1p-r1.
}

The gem5 checkpoints shall be recorded for the following configurations:

\begin{itemize}
\item \textsf{gapbs-big}: \textsf{bfs}, \textsf{cc}, each with 1/16/32 cores with 8\,GB or 16\,GB simulated memory.
\item \textsf{parsec}: \textsf{blackscholes}, \textsf{freqmine}, each with 1/4/16 cores with 3\,GB or 16\,GB simulated memory.
\item \textsf{spec2017}: \textsf{bwaves}, \textsf{gcc}, \textsf{mcf}, \textsf{omnetpp}, each with 1 core with 3\,GB simulated memory.
\end{itemize}

Similarly, kernel and disk files can be organized as:

\dirtree{%
.1 disk-ker/.
.2 \purple{gapbs}/ \DTcomment{kernels and disk images}.
.2 \olive{parsec-3.0}/ \DTcomment{kernels and disk images}.
.2 \teal{spec2017}/ \DTcomment{kernels and disk images}.
}





\subsection{Installation}

Prepare both the \textsf{ckpt/} and \textsf{disk-ker/} directories and place them in a local path with sufficient storage. Then Download the modified gem5 repository containing our profiler (\textsf{gem5-prof}), which has the following structure:

\dirtree{%
.1 gem5-prof/.
.2 ext/.
.3 prof/ \DTcomment{Profiler C++ code. \uline{1, Compile using \texttt{make}}}.
.2 script/ \DTcomment{Various scripts used in the project}.
.3 x86/.
.4 m5out/ \DTcomment{Directory for simulation outputs}.
.4 gapbs/.
.5 ckpt@ \DTcomment{\uline{2, User created.} Softlink to \red{\sffamily gapbs-big}/}.
.5 env.sh \DTcomment{\uline{3, Configure disk and kernel paths to \purple{\sffamily gapbs}/}.}.
.4 parsec-3.0/.
.5 ckpt@ \DTcomment{\uline{4, User created.} Softlink to \orange{\sffamily parsec}/}.
.5 env.sh \DTcomment{\uline{5, Configure disk and kernel paths to \olive{\sffamily parsec-3.0}/}.}.
.4 spec2017/.
.5 ckpt@ \DTcomment{\uline{6, User created.} Softlink to \blue{\sffamily spec2017}/}.
.5 env.sh \DTcomment{\uline{7, Configure disk and kernel paths to \teal{\sffamily spec2017}/}.}.
}

After downloading \textsf{gem5-prof/}, please follow the underlined instructions above to:

\begin{itemize}
    \item Compile the profiler binary (Step 1).
    \item Create the required symbolic links to \textsf{ckpt/*} that point to the corresponding gem5 checkpoints (Step 2, 4, 6). 
    \item Update the environment variables in each \texttt{env.sh} script (Step 3, 5, 7) so that the paths in each \texttt{env.sh} correctly point to the correspoding disk image and kernel binary in \texttt{disk-ker}.
\end{itemize}

\textbf{Note.} One can organize the checkpoints and disk/kernel files in any way that suits one's setup. However, please check the script \texttt{script/x86/run-host-gem5.sh} and ensure that all paths point to the correct places.

\subsection{Experiment workflow}

First compile gem5:

\begin{verbatim}
  cd gem5-prof
  scons build/[build_opt]/gem5.debug [-j X]
\end{verbatim}

All experiments are launched using a Celery-based task framework. Navigate to the experiment launcher:

\begin{verbatim}
  cd gem5-prof/script/x86/cescri
\end{verbatim}

Start the full experiment with:

\begin{verbatim}
  ./kickstart.sh 4 3600
\end{verbatim}

This command:

\begin{itemize}
\item schedules all gem5 tasks over 4 Celery workers
\item runs each simulation with a timeout of 3600 seconds
\end{itemize}

using the experiment configurations located in:

\begin{verbatim}
  gem5-prof/script/x86/cescri/cfg
\end{verbatim}

These configurations specify the benchmark suites, number of cores, CPU models, memory sizes, and CPU microarchitectures. The paper uses \texttt{zen5} by default; however, we also provide three additional microarchitectures---\texttt{lioncove}, \texttt{skymont}, and \texttt{zen4}---located in \texttt{script/x86/arch}. 

The Celery worker log is located at:

\begin{verbatim}
  gem5-prof/script/x86/log
\end{verbatim}

When all tasks complete successfully, the log will show:

\begin{verbatim}
  ==============================
  All gem5 tasks finished!
  ==============================
\end{verbatim}

\subsection{Evaluation and expected results}

After all simulations have finished, plots can be generated using the
provided parsing scripts. Navigate to the parser directory:

\begin{verbatim}
  cd gem5-prof/script/x86/parser
\end{verbatim}

Generate the figures corresponding to the results reported in the paper:

\begin{verbatim}
  ./plot-all-AS.sh
  ./plot-all-TS.sh
  ./plot-all-O3.sh
\end{verbatim}

These scripts process gem5 output data and generate plots corresponding to the evaluation results presented in~\autoref{fig:AS:ae}--\autoref{fig:O3:ae} of the paper. All generated figures are located in:

\begin{verbatim}
  gem5-prof/script/x86/parser/fig/
\end{verbatim}

\subsection{Experiment customization (more than what the paper has)}
\label{ae:g}

The artifact also supports additional analysis of various CPU and Ruby subsystem components (\eg{}, individual O3 pipeline stages) and their interactions. We provide a large number of pre-defined result-exploration scripts (125 in total) beyond those used in the paper. These configurations are located in:

\begin{verbatim}
  gem5-prof/script/x86/parser/cfg
\end{verbatim}

All configurations are defined in Python and provide flexible examination options, including a whitelist (only the runtimes of these functions are shown in a plotted figure), a blacklist (these functions are excluded when calculating gem5 runtime breakdown), and various knobs to control figure plotting. For example, one can specify the root function and whether to fold child functions in a sampled gem5 call-stack: \texttt{level=1} folds child functions, \texttt{level=2} expands children by one level, and \texttt{level=-1} expands all children to the leaf level.

We have put more holistic results in the following arXiv report: Johan S\"oderstr\"om and Yuan Yao, ``Anatomy of the gem5 Simulator: AtomicSimpleCPU, TimingSimpleCPU, O3CPU, and Their Interaction with the Ruby Memory System,'' \textit{arXiv preprint arXiv:2508.18043,} 2025. \url{https://arxiv.org/abs/2508.18043}


\subsection{Methodology}

Submission, reviewing and artifact badging methodology:

\begin{itemize}
  \item \href{https://www.acm.org/publications/policies/artifact-review-and-badging-current}{ACM Artifact Review and Badging Policy}
  \item \url{https://cTuning.org/ae}
\end{itemize}

\end{document}